\documentclass[journal=nalefd,manuscript=article]{achemso}

\usepackage[version=3]{mhchem} 
\usepackage[T1]{fontenc}       
\usepackage{amssymb,amsfonts,amsmath}
\usepackage{xcolor}

\author{Mehrshad Mehboudi}
\affiliation{Department of Physics. University of Arkansas. Fayetteville AR 72701, USA}
\author{Alex M. Dorio}
\affiliation{Department of Physics. University of Arkansas. Fayetteville AR 72701, USA}
\author{Wenjuan Zhu}
\affiliation{Department of Electrical and Computer Engineering. University of Illinois. Urbana IL 61820, USA}
\author{Arend van der Zande}
\affiliation{Department of Mechanical Science and Engineering. University of Illinois. Urbana IL 61820, USA}
\author{Hugh O. H. Churchill}
\affiliation{Department of Physics. University of Arkansas. Fayetteville AR 72701, USA}
\author{Alejandro A. Pacheco-Sanjuan}
\affiliation{Departamento de Ingenier{\'\i}a Mec{\'a}nica. Universidad del Norte. Barranquilla, Colombia}
\author{Edmund O. Harriss}
\affiliation{Department of Mathematical Sciences. University of Arkansas. Fayetteville AR 72701, USA}
\author{Pradeep Kumar}
\affiliation{Department of Physics. University of Arkansas. Fayetteville AR 72701, USA}
\author{Salvador Barraza-Lopez}
\affiliation{Department of Physics. University of Arkansas. Fayetteville AR 72701, USA}
\email{sbarraza@uark.edu}

\title{Two-dimensional disorder in black phosphorus and monochalcogenide monolayers}

\abbreviations{2D,BP,MMs}

\begin{document}



\begin{abstract}
Ridged, orthorhombic two-dimensional atomic crystals with a bulk {\em Pnma} structure such as black phosphorus and monochalcogenide monolayers are an exciting and novel material platform for a host of applications. Key to their crystallinity, monolayers of these materials have a four-fold degenerate structural ground state, and a single energy scale $E_C$ (representing the elastic energy required to switch the longer lattice vector along the $x-$ or $y-$direction) determines how disordered these monolayers are at finite temperature. Disorder arises when nearest neighboring atoms become gently reassigned as the system is thermally excited beyond a critical temperature $T_c$ that is proportional to $E_C/k_B$.  $E_C$ is tunable by chemical composition and it leads to a classification of these materials into two categories: (i) Those for which $E_C\ge k_BT_m$, and (ii) those having $k_BT_m>E_C\ge 0$, where $T_m$ is a given material's melting temperature. Black phosphorus and SiS monolayers belong to category (i): these materials do not display an intermediate order-disorder transition and melt directly. All other monochalcogenide monolayers with $E_C>0$ belonging to class (ii) will undergo a two-dimensional transition prior to melting. $E_C/k_B$ is slightly larger than room temperature for GeS and GeSe, and smaller than 300 K for SnS and SnSe monolayers, so that these materials transition near room temperature. The onset of this generic atomistic phenomena is captured by a planar Potts model up to the order-disorder transition. The order-disorder phase transition in two dimensions described here is at the origin of the {\em Cmcm} phase being discussed within the context of bulk layered SnSe.
\end{abstract}

\section{Introduction}

Monolayers of layered orthorhombic materials \cite{BP1,BP2,BP3,BP4,BP5,lefebre,antunez,LLi,zhang,dravid,zhao,kaxiras,singh,zhu,gomes,gomes2,li,benjamin}
can become disordered at room temperature.

Graphene \cite{RMP,katsnelsonbook} and other 2D atomic materials such as hexagonal boron nitride and transition-metal dichalcogenide monolayers \cite{hBN,MoS2} have a non-degenerate structural ground state that is key to their stability at room temperature. On the other hand, the ridged structure of black phosphorus monolayers and other materials with a similar atomistic structure leads to their celebrated anisotropic electron and optical properties \cite{BP2}. At the same time, such unique atomic arrangement has striking consequences for crystalline order \cite{Onsager,Potts,MerminWagner,KT,NelsonBook,Perk1,Fertig,Perk2,Loverde,Clarke,Han,Glotzer,Yao,Schiffer}
that remain unexplored up to date.

Indeed, the remarkable multifunctionality of ferroelectrics largely originates from the degeneracies of their structural ground state,\cite{Laurent} and degeneracies of the structural ground state lead to well-known mechanical instabilities in two-dimensional critical lattices at finite temperature as well \cite{Mao}. If the structural ground state of black phosphorene (BP), black arsenene, and monochalcogenide monolayers (MMs) turned out to be degenerate, this family of two-dimensional materials must necessarily and inevitably display in-plane disorder at finite temperature. Structural degeneracies may even be key to explain the {\em Pnma-Cmcm} transition seen in bulk samples that has drawn considerable excitement in the thermoelectric community \cite{dravid,delaire}. Structural degeneracies of two-dimensional atomic materials may open the door for new Physics, and may also lead to new and completely unexplored material functionalities that could be controlled with temperature.

\begin{figure}[tb]
\includegraphics[width=0.6\textwidth]{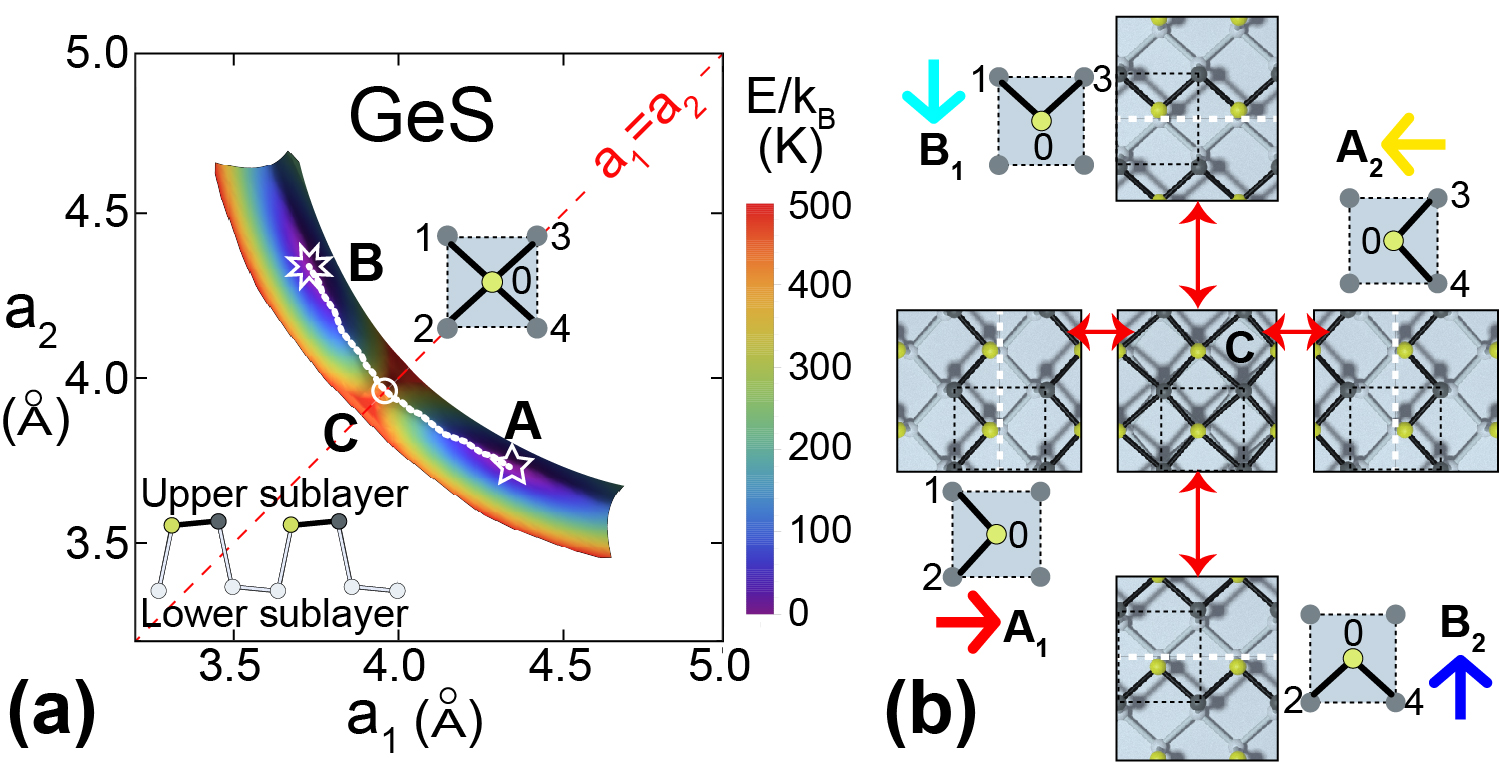}
\caption{(a) The elastic energy landscape $E(a_1,a_2)$ as a function of lattice parameters $a_1$ and $a_2$ is generic to all monolayers with a {\em Pnma} structure, and it is exemplified on a GeS monolayer at zero temperature. A dashed white curve joins points $A$ and $B$ at two degenerate minima ($E_A=E_B=0$). The circle labeled $C$ at $(4.0\AA,4.0\AA)$ is a saddle point with an atomistic structure in which atom 0 forms bonds to four in-plane neighbors, and the elastic energy barrier is defined by $E_C$. (b) Atomistic decorations (i.e., the specific pair of atoms bonding atom 0) increase the structural degeneracy at points $A$ and $B$. The four degenerate ground states are named $A_1$, $A_2$, $B_1$ and $B_2$, and assigned in-plane arrows that label them uniquely.}
\label{fig:fig1}
\end{figure}

This study contains a discussion of degeneracies of the structural ground state of monolayers with a {\em Pnma} structure at zero temperature; the determination of an energy $E_C$ that depends on atomic number and sets the energy scale  for elastic transitions among degenerate ground states; Car-Parrinello molecular dynamics (MD) calculations at finite temperature (carried out on a code that employs localized orbital basis sets) that permit relating the specific heat and an order parameter to $E_C$; and a coarse-grained in-plane (clock) Potts model with $q=4$ that matches the MD data up to the order-disorder transition at a critical temperature proportional to $E_C/k_B$, where $k_B$ is Boltzmann constant. The two-dimensional (2D) order-disorder transition discovered here must have profound consequences for all material properties, including degradation propensity, and it can be experimentally verified by scanning tunneling microscopy \cite{dravid}, temperature-dependent polarized Raman measurements, specific heat measurements, among other methods.

\section{Results and discussion}
\subsection{Structural degeneracies at zero temperature}\label{sec:bistable}
The degeneracy of the structural ground state at zero temperature is generic to all monolayers having a {\em Pnma} structure, and it is discussed within the context of a GeS monolayer (Figure~1) next.

The first source of degeneracy seen on the elastic energy landscape \cite{EL} (Figure~1(a)) stems from the fact that the elastic energy --the total energy at zero temperature as a function of lattice parameters $E(a_1,a_2)$)-- is degenerate upon exchange of $a_1$ and $a_2$: $E(a_1,a_2)=E(a_2,a_1)$. The degeneracy is highlighted with star-like patterns at ground states $A$ (located at (4.34 \AA,3.73 \AA)) and $B$ (at (3.73 \AA,4.34 \AA)) where $E_A=E_B=0$. The second source of degeneracy, illustrated on Figure~1(b), arises from the two mirror-symmetric black zig-zag patterns that can be created by the basis vectors. (The lower sublayer was white-colored on the structural models to make the patterns more evident.)

The four degenerate structural ground states $A_1$, $A_2$, $B_1$ and $B_2$ on Figure~1(b) occur when specific triads of atoms 1-0-2, 3-0-4, 3-0-1, or 4-0-2 create nearest-neighbor bonds, and this in-plane bond structure can be identified with four in-plane arrows: $\rightarrow$, $\leftarrow$, $\downarrow$,$\uparrow$ \cite{Potts}. These atoms form an extra bond to the lower sublayer making the structure three-fold coordinated, but that lower layer reassigns bonds in a similar manner making a discussion of its rearrangement unnecessary.

It is possible to reassign a nearest-neighbor bond  at zero temperature (or to turn arrows by $\pm\pi/2$) by means of the elastic distortion shown by the white dashed curve on Figure~1(a) that converts ground state $A_1$ ($A_2$) --with bonding atoms 1-0-2 (3-0-4)-- onto ground state $B_1$ ($B_2$), where atoms 1-0-3 (4-0-2) bond.

Indeed, the distortion highlighted by the dashed white curve on Figure~1(a) includes the saddle point $C$ where all 4 atoms bond to atom 0, turning the original zig-zag structure onto an {\em unstable Cmcm} ``checkerboard'' structure with an energy cost $E_C$. When the bond to the lower plane is included, this checkerboard structure is five-fold coordinated. The structure must loose some bonds as the elastic deformation along the white dashed path continues from point $C$ to point $B$, being equally likely to turn into decorations $B_1$ or $B_2$ which are both three-fold coordinated. This is how bonds are reassigned at zero temperature. When the lower sublayer (seen in white on Figure 1(b)) is considered, one sees that two chemical bonds are reassigned per unit cell when structure $A_1$ ($A_2$) turns onto structure $B_1$ ($B_2$).

Direct transitions from decoration $A_1$ ($\rightarrow$) to decoration $A_2$ ($\leftarrow$) ($B_1$ ($\downarrow$) onto $B_2$ ($\uparrow$)) are more costly as they require reassigning twice as many bonds: 1-0-2 to 3-0-4 (or 3-0-1 to 4-0-2), or a $\pi-$arrow rotation on Figure~1(b). These transitions can be achieved in two elastic cycles (for instance, from $\rightarrow$ to $\uparrow$, back to $\leftarrow$), and two-ended red arrows on Figure~1(b) indicate most likely transitions among the four structural ground states. The reassignment of nearest-neighbors is at the core of the 2D disorder to be discussed later on.

\subsection{Tuning the elastic energy barrier $E_C$ with atomic number $Z$} The average atomic number $\bar{Z}$ is defined as follows:
\begin{equation}
\bar{Z}=\frac{1}{4}\sum_{i=1}^4Z_i,
\end{equation}
where the sum is over the four atomic elements on a unit cell, each having atomic number $Z_i$ ($i=1,2,3,4$). It will be shown that both $a_1/a_2$ and $E_C$ evolve with $\bar{Z}$ now.

 The magnitude of $E_C$ was determined through stringent calculations with the {\em VASP} code \cite{vasp1,vasp2} whose details are provided as Supporting Information. As indicated previously, Car-Parrinello MD calculations will also be necessary to verify our main claim, and these calculations are prohibitively expensive on any computational code that employs plane-wave sets. For that reason $E_C$ was also computed with the {\em SIESTA} code \cite{siesta}, as that code will permit carrying out MD calculations at the expense of making a choice for the localized basis set in which electronic wavefunctions are to be expanded. Our choice of basis set, described in the Methods section, is such that the magnitude of the lattice constant for BP agrees reasonably well among these two computational tools.

  The values of $a_1/a_2$ and $E_C$ averaged over their magnitude from three different calculations are displayed in Figure 2 and Table 1 (see Methods). Light compounds such as BP or SiS monolayers ($\bar{Z}=15$) have the largest values of $a_1/a_2$ and $E_C$. On the other hand, ultrathin Pb-based monochalcogenides ($\bar{Z}>48$) have a rock-salt structure so that $a_1/a_2=1$ and $E_C=0$. All remaining monochalcogenide monolayers (MMs) have values of $a_1/a_2$ and $E_C$ lying somewhere in between, which implies a vast tunability of $a_1/a_2$ and $E_C$ with atomic number.

Bond covalency is gradually sacrificed with increasing atomic number to favor a higher atomistic coordination and a weaker (i.e., metallic) bonding. In previous work, we showed that group-IV two-dimensional materials turn from a threefold- to a ninefold-coordinated phase with increasing atomic number \cite{Pablo}. In the present case, a threefold-coordinated structure evolves towards a fivefold-coordinated one with increasing $\bar{Z}$.
Being more specific, $a_1/a_2-1$ decays quite rapidly with $\bar{Z}$ and regardless of the numerical approach employed. We fit:
\begin{equation}
\left(\frac{a_1(\bar{Z})}{a_2(\bar{Z})}-1\right)=b\exp\left(-c_1\bar{Z}\right),
\end{equation}
with $b=3.74$ and $c_1=2/16.5$, and display this trendline as the solid curve on Figure 2(a). The decrease of the ratio $a_1/a_2$ with increasing atomic number implies that the energy needed to reach the intermediate state $C$ with $a_1=a_2$ by an elastic distortion is becoming smaller with $\bar{Z}$ too, and using values of $E_C$ from Table 1 one fits:
\begin{eqnarray}\label{eq:eq3}
E_C(\bar{Z})=d\exp\left(-c_2\bar{Z}\right),
\end{eqnarray}
with numerical parameters $d=37,650 K$ and $c_2=1/6$ (this trendline is shown as a straight line in Figure 2(b)).

\begin{figure}[tb]
\includegraphics[width=0.6\textwidth]{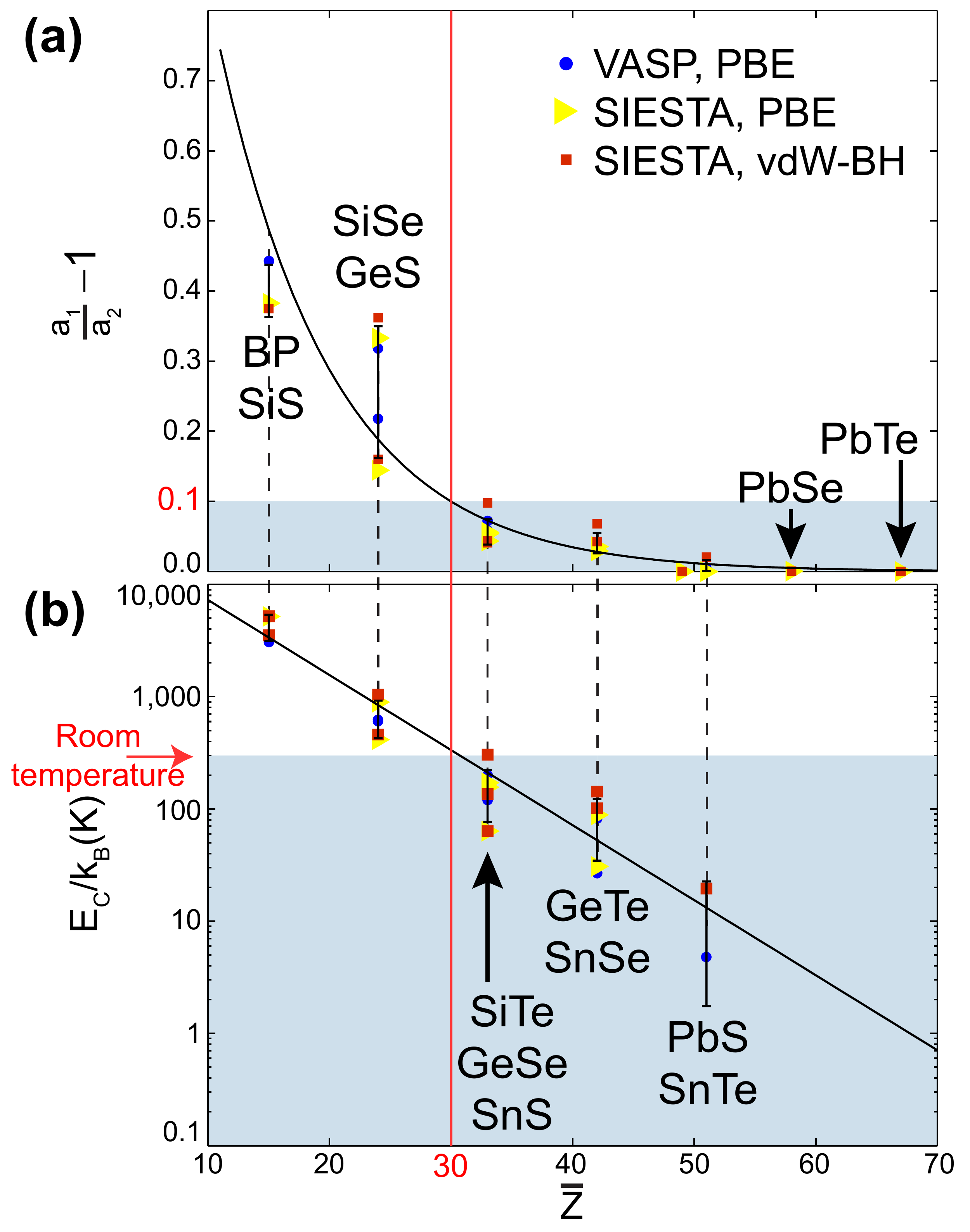}
\caption{(a) The ratio $a_1/a_2$ among orthogonal in-plane lattice constants decreases exponentially with the mean atomic number $\bar{Z}$, and (b) $E_C$ decays exponentially with $\bar{Z}$ as well. $E_C/k_B<300$ K (and $a_1/a_2\leq 1.1$) for $\mathbf{Z}\geq 30$, prompting the question whether three-fold coordinated GeSe, SnS and SnSe monolayers are disordered near room temperature. Structures with $a_1\simeq a_2$ display a five-fold-coordinated and non-degenerate ground state with $E_C\simeq 0$. Solid lines are fits whose parameters are given in the main text. (See Methods.)}
\label{fig:fig2}
\end{figure}

\begin{table}[tb]
\caption{Ratio $a_1/a_2$ among lattice parameters at zero temperature, and  the elastic energy barrier $E_C$ required to switch in between degenerate ground states for BP and twelve MMs. $E_C$ decays exponentially with the average atomic number $\bar{Z}$ (Equation~[2] on the main text). Data scatter arises from the fact that three numerical codes were employed in computing these quantities (see Methods). Experimental melting temperatures of bulk samples are also included for comparison purposes\cite{HCP}.}
\begin{tabular}{c|c|c|c|c}
& & & &Melting temperature\\
Compound& $\bar{Z}$& ${a_1}/{a_2}$ &  $E_C/k_B$ ($K$)  & $T_m$ (bulk) ($K$)\\
\hline
BP & 15 &1.38 $\pm0.01$& 5159	$\pm$75      & 883\\
SiS	    &15&1.40 $\pm0.04$	& 3536	$\pm$462 & 1173\\
SiSe	&24&1.27 $\pm0.11$	& 730	$\pm$446 & --\\
GeS	    &24&1.24$\pm0.09$	&653	$\pm$221 & 888\\
GeSe	&33&1.08$\pm0.02$	&220	$\pm$76  & 940\\
SiTe	&33&1.05$\pm0.01$	& 154	$\pm$24  & --\\
SnS	    &33&1.05$\pm0.01$ & 63	$\pm$0   & 1153\\
GeTe	&42&1.04$\pm0.00$	& 95	$\pm$9   & 998\\
SnSe	&42&1.04$\pm0.02$ &87	$\pm$79      & 1134\\
PbS	    &49&1.00$\pm0.00$ & 0 & 1391\\
SnTe	&51&1.01$\pm0.01$	& 10	$\pm$14  & 1063\\
PbSe	&58&1.00$\pm0.00$ & 0 & 1351\\
PbTe	&67&1.00$\pm0.00$	& 0 & 1197\\
\end{tabular}
\end{table}

Experimental values for the melting temperature $T_m$ of bulk compounds are reported in Table 1. We make an additional point by assuming that the melting temperature of monolayers is relatively close to $T_m$. $T_m$ permits a classification of these two-dimensional materials into two groups: (i) those having $E_C\ge k_BT_m$ and (ii) those where $k_BT_m>E_C\ge 0$.  Black phosphorus and SiS monolayers belong to category (i). Given the error bars in $E_C$, SiSe appears to be borderline between class (i) and (ii). All other MMs belong to class (ii).

The classification introduced in previous paragraph can be used to draw a direct connection to experiments on BP monolayers: Most theory developed for BP has been carried out under the implicit assumption that its atomistic structure does not drastically change in between 0 K and room temperature, and this is confirmed by experiments. Such agreement appears to counter the statement that materials with a {\em Pnma} structure undergo an order-disorder transition. But its rather large magnitude of $E_C$ --higher than its melting temperature-- prevents this material from thermally exploring its degenerate ground states up until it melts and solves the apparent contradiction.

The significance of Equation (3) can be hardly overstated, for it indicates that
$E_C$ is tunable by the choice of compound in Table 1, all the way from 0 K and up to temperatures above $T_m$. In stark contrast to BP monolayers, some MMs undergo a 2D order-disorder transition, so that the implicit assumption that structural symmetries obtained at zero temperature remain at finite temperature does not hold true for many MMs.

\subsection{Two-dimensional disorder at finite temperature}
So far, elastic transitions among $q=4$ degenerate ground states at zero temperature have been studied, and the energy $E_C(\bar{Z})$ required to cycle among these four structures  via elastic strain at zero temperature was established. It will now be proven that the generic existence of four ground states with a finite energy barrier $E_C(\bar{Z})$ to switch among these structures leads to two-dimensional disorder on materials belonging to class (ii) by means of a reassignment of nearest-neighboring bonds at finite temperature.

Previous assertion will be demonstrated  from Car-Parrinello molecular dynamics (MD) calculations \cite{CP,PR,siesta} performed for 1,000 fs at 30, 300, and 1,000 K on periodic supercells containing 576 atoms  with all unit cells initially set to the $A_1$ ($\rightarrow$) decoration. These calculations permit establishing that materials with $E_C> k_BT_m$ belonging to category (i) do not show bond reassignment at any of these temperatures, and that materials belonging to category (ii) having $E_C/k_b<300$ K do show disorder at room temperature.

Nelson makes a point that relevant thermodynamical phenomena can be described without recourse to full atomistic detail: coarse-grained descriptions with effective parameters extracted from atomistic data  increase our intuition of the observed phenomena \cite{NelsonBook}. Following this philosophy, an atomic bond was drawn if interatomic distances lie within 10\% of their value on the structure at 0 K.

\begin{figure*}[tb]
\begin{center}
\includegraphics[width=1.0\textwidth]{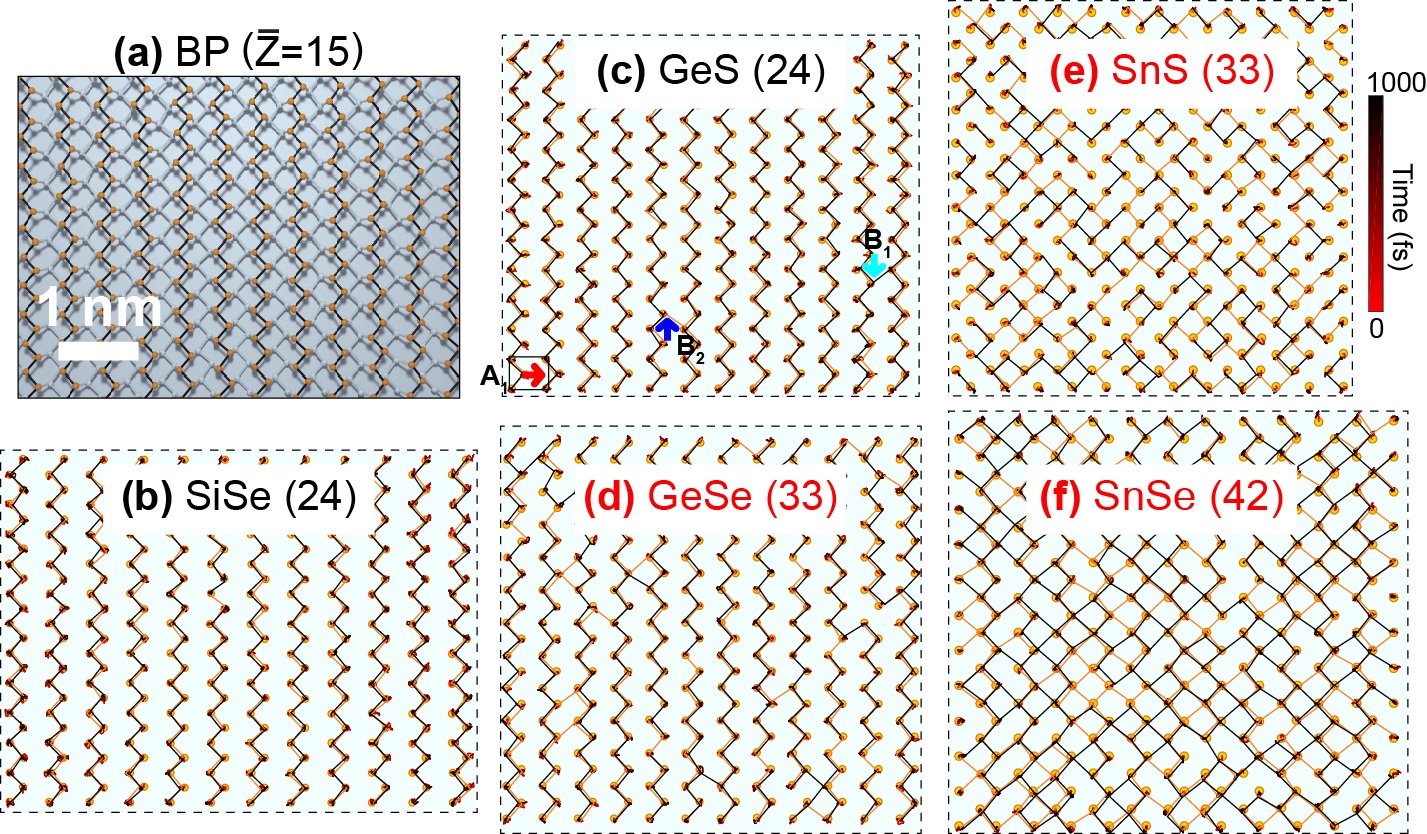}
\caption{(a) Structural snapshot for BP at 300 K. (b-d): MD trajectories at 300 K for atoms in the upper sublayer for SiSe, GeS, GeSe, SnS, and SnSe, respectively. Orange circles depict initial atomic positions, and trajectories during the MD simulation can be seen by continuous curves about the initial positions. Bonds were drawn at 500 and 950 fs to inform of the reassignment of nearest neighbors as the dynamics unfold. Reassignment of nearest neighbors is generic to MMs, and has been explicitly marked by arrows on subplot (c).}
\end{center}
\label{fig:fig3}
\end{figure*}

The total energy equilibrates within 500 fs (Supporting Information), and Figure 3(a) shows a snapshot at the 1000 fs step for a BP monolayer ($\bar{Z}=15$ and $E_C=5159\pm 75$ K) on a periodic 12$\times$12 supercell at 300 K. The vertical zig-zag patterns shown by black bonds consistent with a $A_1$ decoration (Figure 1(b)) that are characteristic of materials with a {\em Pnma} structure can clearly be resolved throughout the dynamical evolution. The zig-zag pattern remains unchanged throughout the entire MD evolution. Figure 3(a) thus demonstrates that a BP monolayer retains its {\em Pnma} structure up to room temperature, as it was hypothesized in previous subsection.

To investigate the hypothesis of two-dimensional disorder created from reassignment of nearest-neighbor bonds on MMs qualitatively, we disclose whether the vertical zig-zag patterns (characteristic of a material with all unit cells set to the $A_1$ decoration) become altered at finite temperature at two time steps, a task simplified by visualizing the upper sublayer only.

We show in Figure 3(b) the time evolution of the atomic positions for atoms belonging to the upper sublayer of SiSe at 300 K ($\bar{Z}=24$ and $E_C=730\pm446$ K). As expected from its large value of $E_C$, the vertical zig-zag pattern can be seen on the figure at all times (the pattern is shown at 500 fs in orange, and 950 fs in black solid lines). The lack of nearest neighbors being reassigned in Figures 3(a) and 3(b) does verify the intuition derived in discussing Figures 1 and 2 in the sense that these materials remain structurally stable at room temperature.

Excitingly, results from MD runs for GeS on Figure 3(c) ($\bar{Z}=24$ and $E_C=653\pm 221$ K and still before the full transition)  begin to display the incipient bond reassignment described in Figure 1(b). A unit cell acquires a $B_2$ ($\uparrow$) decoration at 500 fs, and another unit cell displays a $B_1$ ($\downarrow$) decoration at 950 fs (nearest-neighbors being reassigned evolve as a function of time). These decorations are singled out explicitly by colored arrows. As predicted in Figure 1, these atomic rearrangements are at the onset of the order-disorder transition, and the recommitment of nearest neighbors seen in those two instances induce local strain as they try to stabilize unit cells in which the long axis lines up vertically (the long axis lies horizontally on a $A_1-$decorated structure). Full MD movies for GeS at 30 K, 300 K, and 1000 K where bonds on both planes are drawn are also provided as Supporting Information. We demonstrate in the remainder of Figure 3 the effect of increasing $\bar{Z}$ on the 2D order-disorder transition at room temperature.

GeSe in Figure 3(d) ($\bar{Z}=33$ and $E_C=220\pm76$ K) has a larger number of in-plane nearest neighbor atoms being recommitted, a situation that only aggravates on SnS and SnSe (Figures 3(e) and 3(f), respectively). In Heremans' words, Figures 3(c-f) provide an atomistic view of ``a crystallographic phase transition arising under conditions that lead to a collapse of the two-dimensional crystal structure itself.'' \cite{Heremans1}

\begin{figure*}[tb]
\begin{center}
\includegraphics[width=1.0\textwidth]{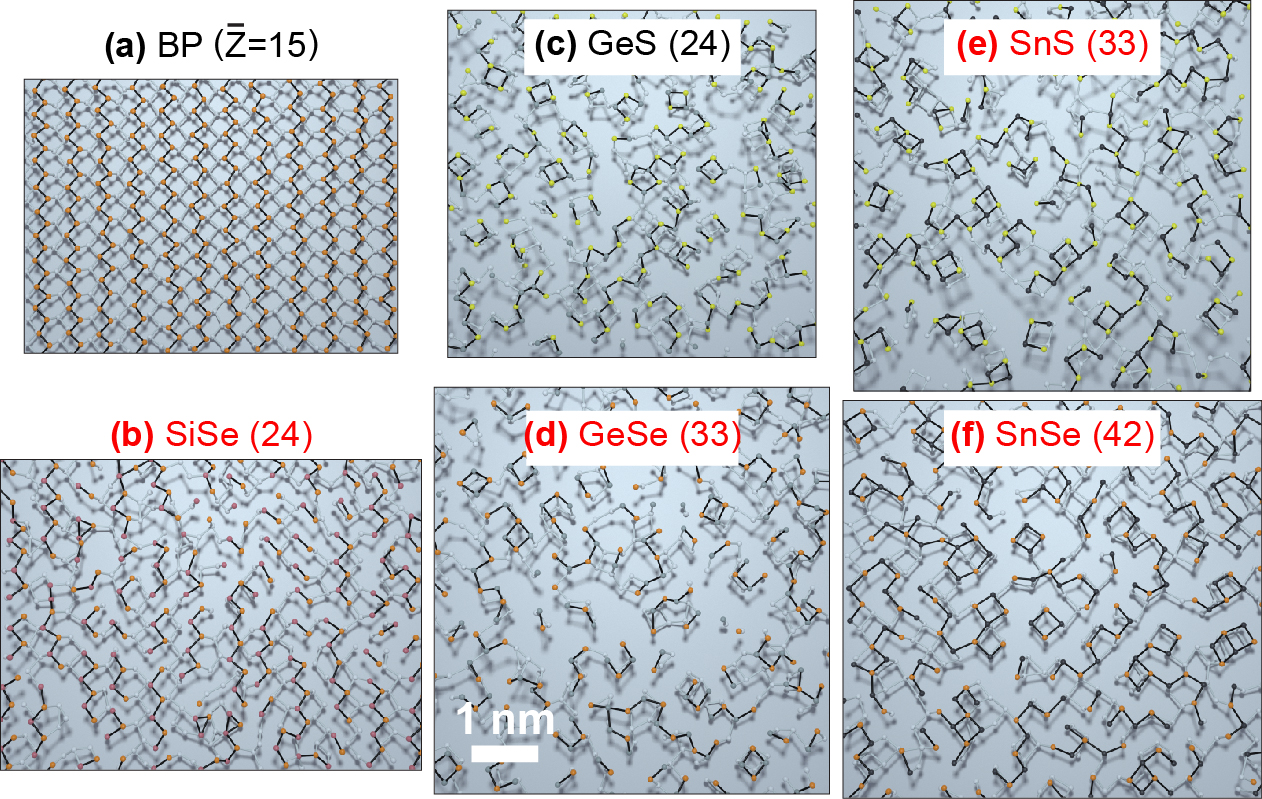}
\caption{Snapshots from Car-Parrinello MD runs at 1,000 K at thermal equilibrium for a BP monolayer and a few MMs. With the notable exception of BP, all other structures are transitioning onto molten phases. The scale bar shown for GeSe applies to all subplots.}
\label{fig:fig4}
\end{center}
\end{figure*}

Snapshots at 1,000 K on Figure~4 indicate that the BP monolayer is still ordered so that it will directly melt at a slightly larger temperature, a finding consistent with its value of $E_C=5,100$ K. All other monolayers are in the process of  transitioning onto molten phases, as indicated by the clustering and dimerization on these images.

 Although illustrative of the atomistic phenomena at play, it is possible to go beyond the visual evidence provided in Figures 3 and 4 in arguing for the two-dimensional transition, and in Table 2 we infer numerically on $12\times 12$ supercells whether the two-dimensional order-disorder transition has occurred or not from the ratio of the average lattice constants $\langle a_1\rangle/\langle a_2\rangle$ at finite temperature (see Methods).

\begin{table*}[tb]
\caption{Evolution of average lattice parameters $\langle a_1\rangle$, $\langle a_2\rangle$ and their ratio {\em versus} temperature ($T$) for 12$\times$12 supercells of BP and eight MMs.}
\small
\begin{tabular}{c||ccc|ccc}
T(K)& $\langle a_1\rangle$(\AA)&$\langle a_2\rangle$(\AA)&$\langle a_1\rangle/\langle a_2\rangle$&$\langle a_1\rangle$(\AA)&$\langle a_2\rangle$(\AA)&$\langle a_1\rangle/\langle a_2\rangle$\\
\hline
\hline
		&&BP&&&SiSe\\
0   	& 4.628$\pm$0.000&	3.365$\pm$0.000&	1.375$\pm$0.000&	5.039$\pm$0.000&	3.728$\pm$0.000&	1.352$\pm$0.000\\
30    & 4.625$\pm$0.050&	3.364$\pm$0.033&	1.375$\pm$0.028&	5.029$\pm$0.058&	3.733$\pm$0.077&	1.347$\pm$0.042\\
300	& 4.628$\pm$0.181&	3.368$\pm$0.103&	1.374$\pm$0.096&	5.054$\pm$0.187&	3.748$\pm$0.272&	1.348$\pm$0.140\\
1000	& 4.632$\pm$0.388&	3.384$\pm$0.207&	1.369$\pm$0.198&	5.090$\pm$1.298&	3.861$\pm$1.302&	1.318$\pm$0.780\\
\hline
\hline
      &&GeS&&&GeSe\\			
0  	 &   4.337$\pm$0.000& 3.739$\pm$0.000&1.160$\pm$0.000&4.473$\pm$0.000&4.074$\pm$0.000&1.098$\pm$0.000\\
30  &  4.338$\pm$0.069& 3.744$\pm$0.044&1.159$\pm$0.032	&4.475$\pm$0.071&4.073	$\pm$0.051&1.099$\pm$0.031\\
300&  4.351$\pm$0.228& 3.764$\pm$0.148&1.156$\pm$0.106	&4.474$\pm$0.266&4.128	$\pm$0.197&1.084$\pm$0.116\\
1000	&4.478$\pm$1.304& 4.271$\pm$0.960&1.048$\pm$0.541&4.942$\pm$1.349&5.019	$\pm$1.386&1.016$\pm$0.541\\
\hline
\hline
&&GeTe&&&SnS\\			
0	    &4.411$\pm$0.000& 4.231$\pm$0.000&1.043$\pm$0.000&4.329$\pm$0.000&4.157	$\pm$0.000&1.041$\pm$0.000\\
30	    &4.414$\pm$0.052& 4.230$\pm$0.055&1.043$\pm$0.026&4.330$\pm$0.061&4.159	$\pm$0.050&1.041$\pm$0.027\\
300  &4.408$\pm$0.199& 4.266$\pm$0.175&1.033$\pm$0.089	&4.258$\pm$0.218&4.245	$\pm$0.190&1.003$\pm$0.096\\
1000	&4.479$\pm$0.720& 4.415$\pm$0.645&1.014$\pm$0.311	&4.837$\pm$1.109&4.704	$\pm$1.171&1.028$\pm$0.492\\
\hline
\hline
&&SnSe&&&SnTe\\		
0		&4.702$\pm$0.000&4.401$\pm$0.000&1.068$\pm$0.000&4.681$\pm$0.000&4.586$\pm$0.000&1.021$\pm$0.000\\
30		&4.695$\pm$0.075&4.406	$\pm$0.060&1.066$\pm$0.032&4.686$\pm$0.060&4.587$\pm$0.057&1.022$\pm$0.026\\
300	&4.687$\pm$0.282&4.468	$\pm$0.254&1.049$\pm$0.123&4.630$\pm$0.219&4.665$\pm$0.220&1.008$\pm$0.094\\
1000	&4.826$\pm$0.595&4.585	$\pm$0.474&1.052$\pm$0.238&4.908$\pm$0.875&4.781$\pm$0.854&1.026	$\pm$0.366\\
\hline
\hline
&&PbS\\				
0		&4.350$\pm$0.000&4.349	$\pm$0.000&1.000$\pm$0.000\\
30		&4.355$\pm$0.049&4.354	$\pm$0.044&1.000$\pm$0.021\\
300	&4.372$\pm$0.179&4.371	$\pm$0.174&1.000$\pm$0.081\\
1000	&4.416$\pm$0.397&4.427	$\pm$0.375&1.002$\pm$0.175
\end{tabular}
\end{table*}

As shown in Figure 3(c), the onset of the transition is dictated by reassignment of nearest neighbors. Reassignment occurs as the atomistic structure explores the four degenerate ground states displayed in Figure 1(b) once temperature reaches a magnitude similar to $E_C/k_B$. Reassignment leads to the macroscopic increase (decrease) of the average local lattice parameter along the vertical (horizontal) direction (see Figure 1(b)) as local strain is released, and hence onto a macroscopic structure in which $\langle a_1\rangle \simeq \langle a_2\rangle$.

BP and SiSe do preserve the magnitude of the ratio $\langle a_1\rangle/\langle a_2\rangle$ to a large extent, which is consistent with their values of $E_C$ equal to 5519$\pm$75 K and 730$\pm$221 K, respectively (Table 1). As seen qualitatively on Figures 3(a) and 3(b), these compounds do preserve their original two-dimensional {\em Pnma} structure until they melt. On the other side of the energy scale set by $E_C$, PbS has $\langle a_1\rangle/\langle a_2\rangle=1$, which is consistent with its magnitude of $E_C=0$ K. All other materials listed in Table 2 are predicted to undergo a 2D order-disorder transition before melting.

Indeed, GeS displays a drastic decrease of $\langle a_1\rangle/\langle a_2\rangle$ in between 300 K and 1000 K (its magnitude drops from $1.156\pm 0.106$ down to $1.048\pm0.541$). This significant drop in between these two temperatures is consistent with a transition temperature of the order of $E_C=653\pm221$ K displayed in Table 1. Analogously, GeSe has a sharp decrease on $\langle a_1\rangle/\langle a_2\rangle$, going from $1.084\pm 0.116$ at 300 K, down to $1.016\pm0.541$ at 1000 K. This also means that a two-dimensional order-disorder transition is occurring roughly after 300 K. Table 1 predicts a transition temperature close to 300 K ($E_C=220\pm 76$ K). We will be able to determine the exact transition temperature for GeS and GeSe later on.

Additionally, SnS has $E_C=63\pm 0$ K for SnS, which implies that a significant drop on  $\langle a_1\rangle/\langle a_2\rangle$ must occur in between 30 K and 300 K. Accordingly, Table 2 points to a decrease from $1.041\pm 0.021$ down to $1.003\pm 0.096$ in between those temperatures.

According to Table 1, GeTe and SnSe should transition at temperatures in between 30 K and 300 K. Both compounds do show a decrease of $\langle a_1\rangle/\langle a_2\rangle$ in between 30 K and 300 K on the second significant digit, but that decrease is not as drastic as registered with other compounds. Each calculation leading to a datapoint in Table 2 consumed about one month of uninterrupted computing, and it appears as if these two compounds may still require additional MD steps to reach a smaller ratio. As far as we know, the MD tool we employ does not offer a numerically sound way to restart a MD calculation. Nevertheless, degeneracies of the ground state lead to mechanical instability at finite temperature \cite{Mao}, and we cannot think of a physical mechanism that will alter the assumption that GeTe and SnSe will continue to reduce their ratio $\langle a_1\rangle/\langle a_2\rangle$ on MD runs performed over longer periods of time. The Potts model, to be introduced later on, will also help us in bringing all numerical results onto a coarse model that describes all the phenomena observed in our MD calculations.

SnTe presents a different kind of numerical challenges. This material has a ratio of $\langle a_1\rangle/\langle a_2\rangle$=1.02 at zero K and a barrier $E_C$ roughly larger than 1 meV. $E_C$ is (roughly) the actual energy difference per unit cell between the ordered and disordered phases one is attempting to resolve. This numerical consideration makes it increasingly challenging to do better on this material: its magnitude of $E_C$ is minuscule in comparison with the other materials listed in Table 1.

A final physical consideration must be added to the discussion of SnS, SnSe, SnTe, and PbS on Table 2: lattice constants loose their meaning on a molten (i.e., non-crystalline) phase, so the increase on $\langle a_1\rangle/\langle a_2\rangle$  at 1000 K with respect to their magnitude reported at 300 K on these compounds should not be given a heavy weight.

Table 2 continues to build the evidence towards a generic 2D order-disorder transition, and we next report the transition temperature $T_c$ for GeS and GeSe through a study of the evolution of energetics and order parameters with an increased temperature resolution.

\subsection{Energetics and order parameter as a function of temperature}

\begin{figure*}[tb]
\begin{center}
\includegraphics[width=.6\textwidth]{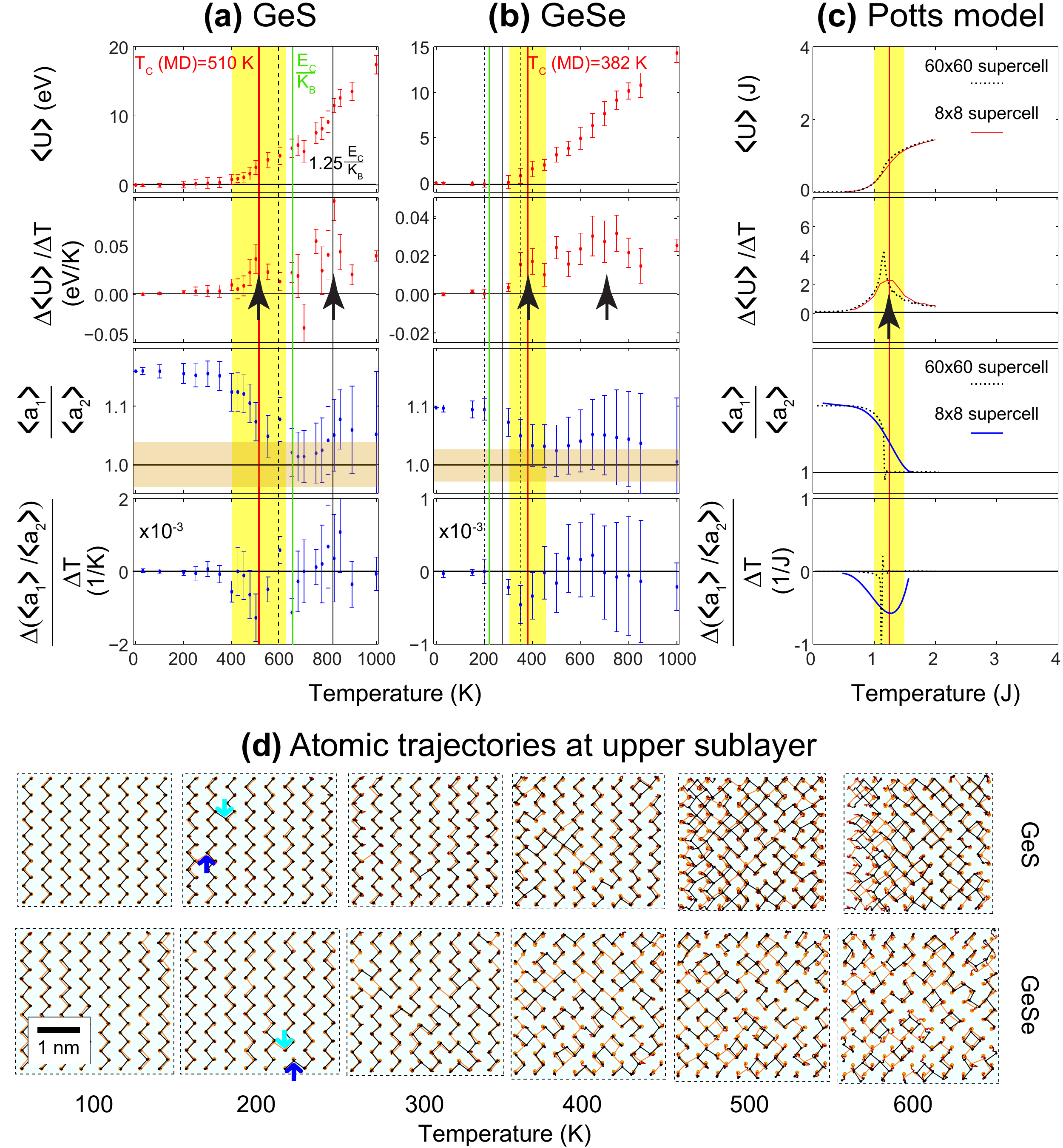}
\caption{Configurational energy $\langle U\rangle$, its thermal derivative, $\langle a_1\rangle/\langle a_2\rangle$, and its thermal derivative for (a) GeS and (b) GeSe. (c) These MD results can be explained up to the two-dimensional order-disorder transition with a $q=4$ Potts model, whose predictions for a 8$\times$8 (60$\times$60) supercell are shown by solid (dashed) curves. (d) Atomistic trajectories for GeS and GeSe show the onset of the transition and the disordered structures at higher T. (Actual standard deviations for $\Delta \langle U\rangle/\Delta T$ and $\langle a_1\rangle/\langle a_2\rangle$ are five times larger than those seen on these subplots, and 20 times larger for the $\Delta(\langle a_1\rangle/\langle a_2\rangle)/\Delta T$ subplot.)}
\label{fig:fig5}
\end{center}
\end{figure*}

In Figures 5(a) and 5(b) we address the thermal evolution of the energy and the order parameter for GeS and GeSe with increased resolution. To proceed with celerity, these MD calculations were performed on periodic 8$\times$8 supercells. The total energy, averaged from 500 to 1,000 fs at a given temperature is labeled $\langle E(T)\rangle$, and it has kinetic and configurational (i.e., potential) energy contributions. According to the equipartition theorem, the kinetic energy is proportional to temperature, with $c$ the proportionality constant. This way, the configurational energy $\langle U (T)\rangle$ displayed for GeS and GeSe in Figures~5(a) and 5(b) is $\langle U(T)\rangle=\langle E(T)\rangle -cT$. The similar trends on $\langle U (T)\rangle$ and its thermal derivative obtained for two different MMs provides yet additional proof of the generality of the phenomena shown in Figures 1 and 2 that leads to the two-dimensional order-disorder transition. In a manner consistent with Figure 3, atomistic trajectories from MD calculations are given in Figure~5(d) at many temperatures for GeS and GeSe. Incipient reassignments of nearest neighbors, at the onset of the order-disorder transition, are shown as well.

Phase transitions lead to drastic changes in the specific heat $C_p\equiv d\langle  E (T)\rangle/d T=d\langle  U (T)\rangle/d T+c$ at zero pressure \cite{Huang}. $\langle U (T)\rangle$ is a monotonically increasing function of temperature and it has two marked changes of slope that lead to two peaks on the finite-difference temperature derivative $\Delta \langle  U (T)\rangle/\Delta T$ in Figures~5(a) and 5(b). These two peaks correspond to phase transitions from a crystalline onto a disordered two-dimensional phase, and from the disordered two-dimensional phase onto a molten phase, respectively. The focus of this manuscript is on the 2D order-disorder transition. According to Figures 5(a) and 5(b), the transition temperature $T_c$ is 510 K for GeS ($\bar{Z}=24$), and at 382 K for GeSe ($\bar{Z}=33$). There is a proportionality among $T_c$ obtained from MD calculations (red vertical lines in Figure 5) and $E_C/K_B$ from Table 1 (shown in a green vertical lines in Figures 5(a) and 5(b)), so that Car-Parrinello MD results do validate the basic intuition drawn from Figures 1 and 2 within the limitations in numerical precision from these hero-type, proof-of-concept MD calculations.

The order parameter $\langle a_1\rangle/\langle a_2\rangle$ for GeS and GeSe transitions onto a value closer to unity at $T_c$, a fact that can be better appreciated in the numerical derivative plots $\Delta(\langle a_1\rangle/\langle a_2\rangle)/\Delta T$ shown as the lowermost subplots in Figures 5(a) and 5(b). Figure 5 reproduces the magnitudes of this ratio reported in Table 2 for larger supercells. Figures 5(a) and 5(b) continue to provide further numerical validation to the hypotheses set forth in discussing Figures 1 and 2 and demonstrate in an even more convincingly manner the unavoidable disordered nature of MMs at finite temperature.

The two-dimensional order-disorder transition is quite relevant for practical room-temperature applications based on MMs, and a model that describes it is provided next.

\subsection{2D order-disorder transition and Potts model}

Starting with a crystalline structure in which all unit cells have an $A_1$ ($\rightarrow$) decoration, a local energy penalty equal to $J$ ($=E_C$) is given to a neighboring unit cell that reassigns two bonds in acquiring the $B_1$ ($\downarrow$) or the $B_2$ ($\uparrow$) decoration. A direct transition from a $A_1$ ($\rightarrow$) to a $A_2$ ($\leftarrow$) decoration requires reassigning four bonds, and is thus given an energy penalty equal to $2J$. This prescription leads to a 2D (clock) Potts model with $q=4$ states \cite{Potts} whose dynamical behavior is characterized by an in-plane spin Hamiltonian on a square lattice with the following nearest-neighbor coupling:
\begin{equation}
\hat{H}=-J\sum_{\langle i,j\rangle}\cos(\theta_i-\theta_j).
\end{equation}
 In previous equation, $\theta_i$  and $\theta_j$ can take any of the four ($q=4$) values 0 ($\rightarrow$), $\pi/2$ ($\uparrow$), $\pi$ ($\leftarrow$), or $3\pi/2$ ($\downarrow$). At low temperatures, the transition with energy penalty $J$ becomes dominant, and the system behaves as if there was just a single coupling. Figure~5(c) displays the predictions from the model, which reproduces the temperature dependence of the configurational energy $\langle U\rangle$ and $\langle a_1\rangle/\langle a_2\rangle$ from Car-Parrinello MD, a remarkable agreement in light of the simplicity of the model and the complexity of the {\em ab-initio} calculations: Potts model provides an appealingly simple physical picture for the 2D order-disorder transition.

\begin{figure}[tb]
\begin{center}
\includegraphics[width=0.6\textwidth]{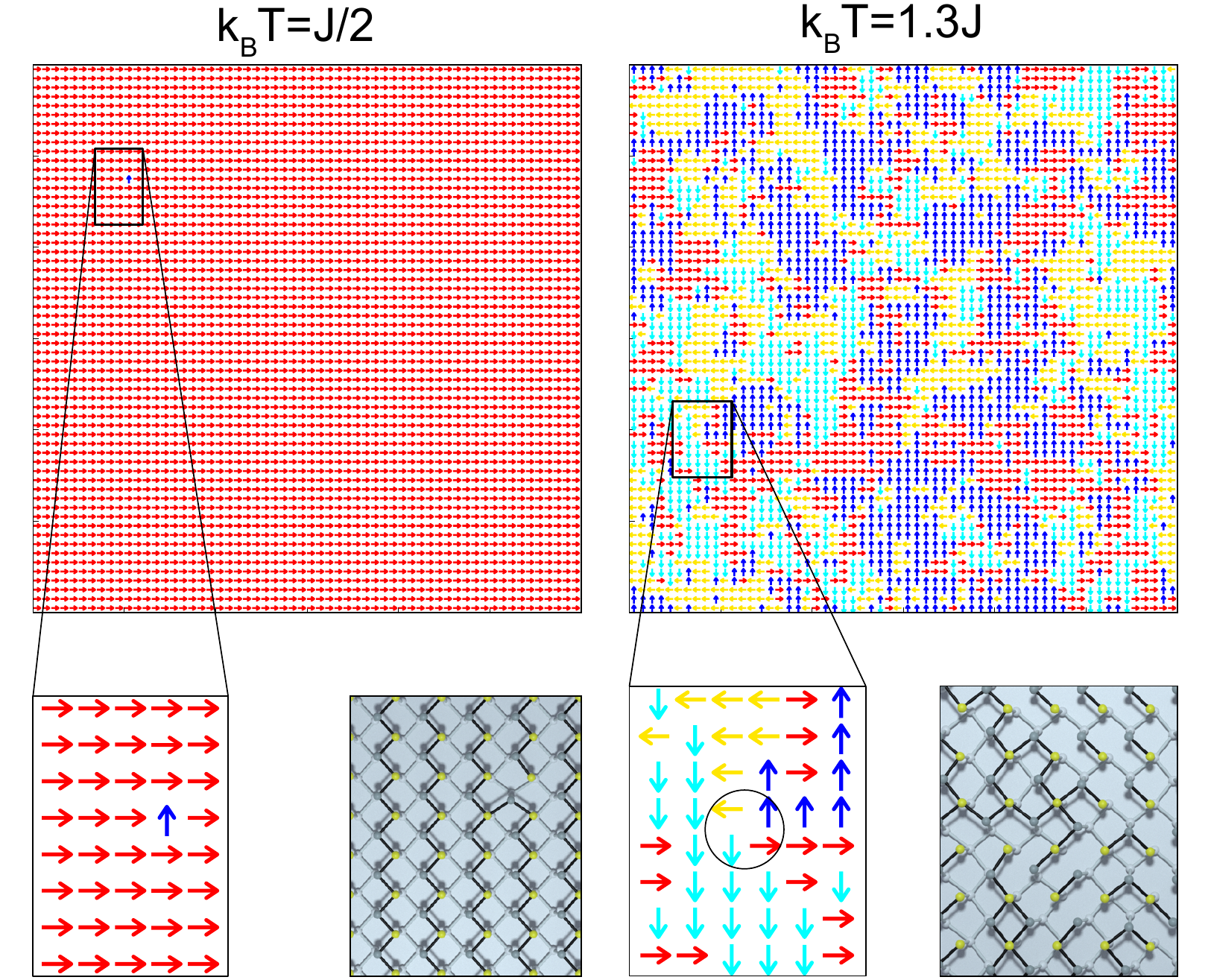}
\caption{The Potts model is employed to ascertain domain sizes at experimentally-relevant scales. A vortex structure seen on the disordered state has been highlighted as a zoom-in, and unrelaxed atomistic models are attached to highlight the inherent atomistic disorder.}
\label{fig:fig6}
\end{center}
\end{figure}

It is impossible to sample larger cells with the {\em ab-initio} method, so Potts model was employed to increase sampling statistics which leads to domain sizes shown in Figure~6 on supercells containing $60\times 60$ unit cells, corresponding to an area of $30 \times 30$ nm$^2$. At $T=1.3J/k_B$ it is highly unlikely to see domains with arrows pointing in opposite directions along the same direction due to the higher energy cost required to create such configuration, and intriguing `vortex' configurations can be observed as well.

Structural disorder implies the existence of electronic disorder. Indeed, the electronic wavefunctions below and above the Fermi energy seen in Figure~7 for disordered 2D SnSe exhibit drastic changes at room temperature when contrasted with their appearance in a crystalline structure.

\begin{figure*}[tb]
\begin{center}
\includegraphics[width=0.9\textwidth]{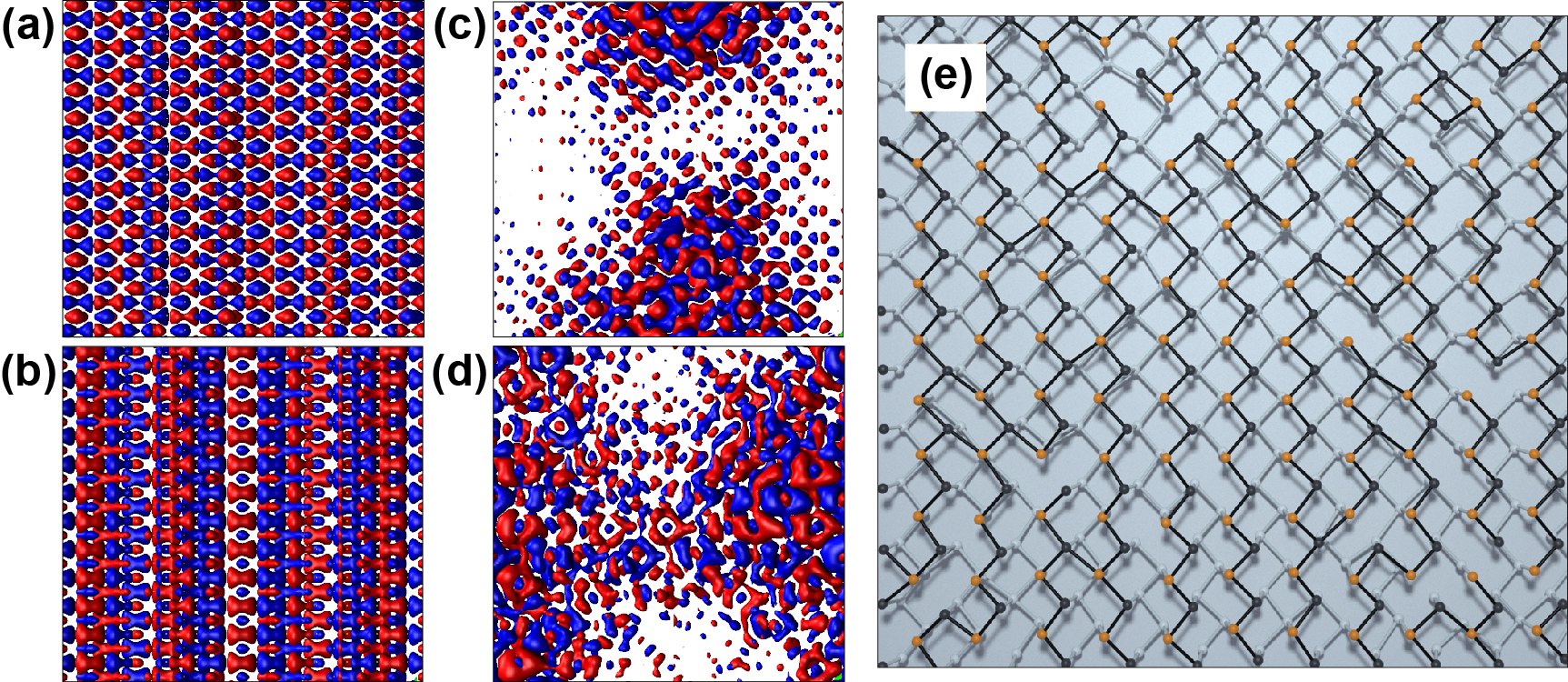}
\caption{Structural disorder creates electronic disorder: The states (a) below and (b) above the Fermi level for SnSe at 0 K. State (c) below and (d) above the Fermi level for SnSe corresponding to the snapshot at 300 K shown in subplot (e).}
\end{center}
\label{fig:fig4}
\end{figure*}

Valleytronics, piezoelectricity, and shift photovoltaics applications \cite{gomes2,li,benjamin} have been predicted for fully crystalline phases; the results provided here thus constrain the choice of materials for these applications near room temperature. At a fundamental level, it will be interesting to explore how these predictions become altered in the presence of 2D disorder, something that is the subject of forthcoming work.

\section{Conclusion}
In summary, the ridged structure of black phosphorus and of layered monochalcogenides leads to 2D disorder as in-plane nearest-neighbors become reassigned at finite temperature. A second transition from this disordered 2D lattice onto a gas phase occurs at a larger temperature, with the remarkable exception of a BP monolayer which has no visible signs of breakdown at 1,000 K so that it will directly melt from the 2D crystalline structure.  Transition temperatures equal to 510 and 382 K were determined for GeS and GeSe, respectively. The phenomena was demonstrated with state-of-the-art large-scale molecular dynamics calculations on periodic supercells, with all relevant interactions computed at the {\em ab initio} level.

An analogy among bond reassignment and four orientations of an arrow leads to a minimal description of the observed phenomena by a Potts model that has $J = E_C$ as its single parameter. This model describes the full-scale energetics through the 2D order-disorder transition remarkably well.

The generic results presented here call to investigate the properties recently predicted on crystalline samples (valleytronics, shift-current photovoltaics, and piezoelectronics) \cite{gomes,gomes2,li,benjamin} as 2D disorder sets in; something that will be pursued in forthcoming work. Having $T_c=510$ K, GeS monolayers appear as the proper crystalline material platform --already available in the bulk form-- for the pursuit of MM-based applications that require crystallinity of the 2D lattice at room temperature. These results apply to freestanding monolayers, and it may be possible to raise the magnitude of the transition temperature $T_c$, through thermal coupling to a substrate or in the bulk. 

The predictions contained here illustrate classic results from the theory of phase transitions and disorder in two-dimensions at work in this novel family of 2D atomic materials.

\section{Methods}
Density-functional theory calculations with the Perdew-Burke-Ernzerhof (PBE) \cite{PBE} and the van der Waals Berland-Per Hyldgaard (BH) \cite{BH} exchange-correlation functionals were carried out on unit cells containing four atoms to determine the elastic energy barrier $E_C$ in Table 1, with the {\em VASP} \cite{vasp1,vasp2} (PBE, Method 1) and {\em SIESTA} \cite{siesta} codes (PBE, Method 2; and BH, Method 3). This required a dedicated deployment of pseudopotentials for the {\em SIESTA} code for most of the chemical elements employed in this work \cite{PseudosPaper}. According to Reference \cite{BH} the BH pseudopotential produces accurate lattice constants and bulk moduli of layered materials and tightly bound solids.

All calculations on single unit cells leading to $E_C$ on Table 1 were performed with a 18$\times$18$\times$1 k-point sampling. {\em SIESTA} calculations had a large Mesh cutoff (used for computing the potential energy on a real-space grid) of 300 Ry. In addition, the electronic density was converged down to 5$\times 10^{-6}$.

Car-Parrinello MD calculations with Method 3 at constant temperature and zero pressure were carried out with the {\em SIESTA} code at temperatures of 30, 300, and 1,000 K for one thousand steps (with a 1fs time step) on periodic supercells containing 576 atoms. These calculations employed a single k-point, a reduced mesh cutoff of 200 Ry, and a reduced tolerance on the electronic cycle of 10$^{-4}$, and require about a full month to end for a given temperature, running on 128 processors.

A more detailed analysis of the total energy as a function of temperature for GeS and GeSe monolayers was performed on smaller periodic $8\times 8$ supercells containing 256 atoms, still employing a single k-point, a reduced mesh cutoff of 200 Ry, and a reduced tolerance on the electronic cycle of 10$^{-4}$. Calculation of an individual datapoint on Figs. 5(a-b) accrues two weeks running on 128 processors.

The average lattice constants $\langle a_1\rangle$ and $\langle a_2\rangle$ were obtained for nine compounds from the distances among each of the four atoms on a given unit cell with respect to the position of said basis atoms in neighboring unit cells once thermal equilibrium was reached, leading to an average over 12$\times$12$\times$500 individual values from {\em SIESTA} calculations with BH pseudopotentials that is reported up to three significant digits. By definition $\langle a_1\rangle=a_1$ and $\langle a_2\rangle=a_2$ on the crystalline structures at 0 K, whose values were then obtained from single unit cell calculations. Values of $a_1$ and $a_2$ at 0 K are consistent with these shown in Table 1, which were in turn obtained as averages at zero temperature over the three different computational methods. $\langle a_1\rangle$ and $\langle a_2\rangle$ are also indicators of the numerical precision being achieved in our MD calculations.

All {\em SIESTA} calculations reported on this paper were performed using standard basis sets with an energy shift of 0.01 eV; this choice was made because that the lattice constants for BP agree among {\em SIESTA} and {\em VASP} reasonably well: (4.635, 3.302), (4.625,3.345), and (4.627, 3.365) \AA{} from {\em VASP} with PBE pseudopotentials, {\em SIESTA} with PBE pseudopotentials, and {\em SIESTA} with Berland-Per Hyldgaard pseudopotential calculations.

Monte Carlo calculations of the configurational energy and the order parameter as predicted from the Potts model were carried on supercells of increasing size with an in-house algorithm.

\begin{acknowledgement}

Conversations with Y. Yang, L. Bellaiche, B.M. Fregoso, L. Shulemburger and F. de Juan are acknowledged. A.M. Dorio acknowledges support from NSF through REU grant DMR-1460754. Molecular dynamics calculations were performed at Arkansas' {\em Trestles} and at TACC's Stampede (Grant  XSEDE TG-PHY090002), and S.B.-L. thanks J. Pummill, D. Chaffin, and P. Wolinski at Arkansas' High-performance computing center for their support. The authors declare no competing financial interest.
\end{acknowledgement}

\begin{suppinfo}
MD movies for GeS at 30K, 300K and 1,000K, details of the calculation of $E_C$ for BP, and a demonstration of thermal equilibration achieved after 500 fs are provided as Supporting Information.

\end{suppinfo}



\providecommand{\latin}[1]{#1}
\providecommand*\mcitethebibliography{\thebibliography}
\csname @ifundefined\endcsname{endmcitethebibliography}
  {\let\endmcitethebibliography\endthebibliography}{}

\end{document}